\documentclass{ieeeaccess}
\usepackage{cite}
\usepackage{amsmath,amssymb,amsfonts}
\usepackage{graphicx}
\usepackage{textcomp}
\usepackage[ruled]{algorithm2e}
\usepackage{kotex}
\usepackage{eqnarray}
\usepackage{cite}
\usepackage{mleftright}

\def\BibTeX{{\rm B\kern-.05em{\sc i\kern-.025em b}\kern-.08em
    T\kern-.1667em\lower.7ex\hbox{E}\kern-.125emX}}
    
\begin{document}

\history{Date of publication xxxx 00, 0000, date of current version xxxx 00, 0000.}
\doi{10.1109/ACCESS.2017.DOI}

\title{Autonomous Power Allocation based on Distributed Deep Learning for Device-to-Device Communication Underlaying Cellular Network}
\author{Jeehyeong~Kim, Joohan~Park, Jaewon~Noh, Sunghyun~Cho}
\address{Department of Computer Science and Engineering, Hanyang University, Korea (e-mail: \{manje111, 1994pjh, wodnjs1451, chopro\}@hanyang.ac.kr)}

\tfootnote{This work was supported by the research fund of Signal Intelligence Research Center supervised by the Defense Acquisition Program Administration and Agency for Defense Development of Korea.}

\markboth
{Author \headeretal: Preparation of Papers for IEEE TRANSACTIONS and JOURNALS}
{Author \headeretal: Preparation of Papers for IEEE TRANSACTIONS and JOURNALS}

\corresp{Corresponding author: Sunghyun Cho (e-mail: chopro@hanyang.ac.kr).}

\begin{abstract}
		    For Device-to-device (D2D) communication of Internet-of-Things (IoT) enabled 5G system, there is a limit to allocating resources considering a complicated interference between different links in a centralized manner. If D2D link is controlled by an enhanced node base station (eNB), and thus, remains a burden on the eNB and it causes delayed latency. This paper proposes a fully autonomous power allocation method for IoT-D2D communication underlaying cellular networks using deep learning. In the proposed scheme, an IoT-D2D transmitter decides the transmit power independently from an eNB and other IoT-D2D devices. In addition, the power set can be nearly optimized by deep learning with distributed manner to achieve higher cell throughput. We present a distributed deep learning architecture in which the devices are trained as a group but operate independently. The deep learning can attain near optimal cell throughput while suppressing interference to eNB. 
	    \end{abstract}


\titlepgskip=-15pt

\maketitle

\section{Introduction}
\label{sec:introduction}
\PARstart{D}{evice} to device (D2D) communication is an emerging technique to able to cope with the increasing mobile traffic demands~\cite{ref:traffic1}. Specifically, Internet of Things (IoT) enabled 5G system is one of the most important system to use D2D communication. Major scenarios of the IoT enabled 5G include remote control or broadcasting alert message by distributed wireless sensors~\cite{ref:iot1, ref:iot2, ref:iot3}. Conventionally, interference management between two links is mainly focused on the D2D communications underlaying cellular system~\cite{ref:D2Dsurvey1, ref:D2Dsurvey2, ref:D2Dsurvey3,ref:D2Dsurvey4}. However, more challenges are still remained in the IoT-D2D enabled 5G system. First of all, the data and control planes would be separated and there are additional small base stations that support only the data plane in the 5G~\cite{ref:D2D_5G}. It means that the base station has to control devices which are covered by multiple small cells. Consequently, the control burden for the base station would be cumulated. In addition, many IoT devices will be deployed with cellular support. If D2D communication supports offloading only in a data plane, the performance of offloading is significantly reduced because of the management overhead to control the D2D connectivity in 5G. The second challenge is the latency. Ultra-low latency is one of the primary requirements of 5G~\cite{ref:D2Dsurvey3}. The time required for resource allocation is one of the major causes of increased latency. The time to request and receive scheduling information from a central node is inevitable in the conventional D2D communications. The conventional D2D communication also has the problem that channel information for all D2D links is required for efficient resource allocation. If all IoT-D2D devices report their channel information periodically, it might be significant burden to control channel and a central node. In addition, the computational overhead in a central node cannot be ignored.

	Therefore, we propose an autonomous power allocation scheme for IoT-D2D devices without involvement of a central node. The proposed scheme operates similarly with a static transmit power decision but it avoids interference between a cellular link and an IoT-D2D link. In addition, to exchange channel information between D2D devices is not required in the proposed scheme because it can operate independently on each D2D device. The proposed autonomous power allocation scheme can maximize the total throughput of D2D links while suppress the interference to cellular networks below a predetermined level. \par

    To achieve these goals, the proposed scheme uses deep learning. Deep learning and deep reinforcement learning have been exploited in various fields of wireless communications and networks~\cite{ref:drl_survey}. Applying deep learning to IoT-based communication is also activated with various approaches~\cite{ref:ex_dl_iot1, ref:ex_dl_iot2, ref:ex_dl_iot3,ref:ex_dl_iot4}. These researches prove that the IoT network is also one of the good candidates to apply deep learning to optimize the performance. Note that the inference requires less computation compared to training and the evolution of IoT hardware is very fast. In addition, on-chip execution with a pre-trained model has proven to be fully feasible~\cite{ref:chip}. Also, the authors in~\cite{ref:light} suggest an extremely efficient deep learning for mobile devices. Both technologies allow deep learning to be used in IoT devices. \par
    
    
    In the proposed scheme, all devices have pre-trained deep learning model to maximize total throughput of D2D links by distributed power allocation. The deep learning model performs the role of sophisticated mapping between local information and global objective function. The proposed scheme based on deep learning has three main features: the distributed decision model which can maximize total cell throughput, the reduced process which uses only location information to eliminate exchanging channel information step, and the customized objective function for deep learning while maintaining interference constraints. The proposed scheme also suggests the methodology to customize an objective function for deep learning. Thus, the proposed scheme can be easily extended to consider other constraints such as energy efficiency. \par

	 The main contributions of this paper are as follows:

	\begin{enumerate}
		\item We propose a power allocation scheme for IoT-D2D communication using only location information. We note that the channel model can be statistically expressed as a function of distance. We propose a learning architecture which implies the overall deriving process including channel model in hidden layers. 
		\item We suggest an autonomous power decision scheme with local information to meet a near-optima. We design a distributed learning architecture for deep learning. We use one deep learning model to train with big data generated by simulation. After training, every IoT-D2D device has the same trained model. It enhances the feasibility of the implementation of the proposed scheme.
		\item We design a customized cost function to optimize an objective function with several constraints in Lagrange multiplier method. It is important that the objective function and constraints have similar scales for deep learning. We design constraints to similar form to the objective function. Then, it is shown that it works well with Lagrange multipliers which can be roughly found.
		\item Consequently, we reduce the time for power decision in D2D communication with two factors: shortening the process by making it autonomously possible and reducing computational complexity. In deep learning, the training process requires a lot of computation and longer processing time compared to the inference.

	\end{enumerate}
	The rest of this paper is organized as follows. In section II, we introduce related studies on IoT-D2D communication. In section III, the proposed method is described in three aspects: a distributed architecture, cost design for learning, and a deep learning process. Section IV presents the results of the actual implementation of the proposed scheme. We show various expressions of the results, including the power distribution of the cells. Finally, the significance of the proposed method is summarized in the conclusion section.

	\section{Related works}
	    \subsection{D2D communications}
	        Many of the D2D communication researches consider IoT enabled 5G system. There are some studies about that D2D communication and small base station are coexisted~\cite{ref:rel1, ref:rel2, ref:karim}. The authors of~\cite{ref:rel1} proposed a graphical solution to obtain an optimal transmission power of reusing nodes and proposed a potential game to solve the radio resource allocation problem in a distributed manner. The authors of~\cite{ref:rel2} used the Stackelberg game to solve the power allocation problem of D2D nodes. The cellular user equipment is considered as the leader of the game and D2D transmitter and small user equipment are considered as the follower of the game. After the setting of leader and followers, they analyze the strategies of leader and follower to obtain the optimal performance. The authors of~\cite{ref:rel3} consider the game theory with incomplete information mechanism. They proposed a static game for resource allocation in multi cell scenario and a repeated game extended from the static game with incomplete information. In addition to comparing and improving SINR performance, there are studies to improve other metrics like energy efficiency or fairness in D2D enabled environment~\cite{ref:rel4, ref:rel5, ref:rel6}. The authors of~\cite{ref:rel4} proposed two heuristic algorithms to allocate resources to cellular and D2D links. In this study, fairness is considered significantly among all the nodes. The cognitive radio situation is considered in~\cite{ref:rel5}. The traditional cellular communication is considered as the primary link and D2D communication is considered as a secondary link. To obtain optimal energy efficiency of the secondary link by protecting minimum rate constraint of primary, the authors proposed an algorithm by considering two transmit covariance matrices of the secondary link. An energy harvesting enabled D2D network is considered in~\cite{ref:rel6}. The optimization problem in this paper has a constraint related to energy harvesting and optimize the rate of the D2D links. For distributed resource allocation in D2D networks, the authors of~\cite{ref:autol} formulated the problem as a stochastic non-cooperative game with multi-agent Q-learning. However, it requires several iterations to converge for each resource allocation. In~\cite{ref:d2d_conv}, a distributed resource allocation scheme was proposed also by using game theory, but it requires additional information exchange.

    \subsection{D2D communications for low latency} 
    The ultra-low latency is the key requirement of the proximity communications~\cite{ref:D2Dsurvey3, ref:D2D_5G}. PC5 interface is considered to support proximity communications in the 3GPP standards~\cite{ref:rel7}. Mode 4 of PC5 interface is considered as a mechanism to allocate radio resource of D2D nodes in a distributed manner. When a node wants to establish a D2D link without the cellular network control, the node uses the PC5 mode 4 with configuration parameters. However, the scheme is not currently specified in the standard, so it needs to be more studied~\cite{ref:rel8}. Therefore, there are many papers to find optimal radio resource allocation mechanisms in various D2D communication scenarios~\cite{ref:rel9, ref:rel10,ref:rel10-1, ref:rel11, ref:rel12}. The binary search algorithm is used to propose an algorithm to guarantee latency and reliability of proximity communications in~\cite{ref:rel9}. In this paper, the authors also proposed a technique of converting the latency constraint into equivalent rate constraint to solve an optimization problem easily. The situation that IEEE 802.11p protocol and LTE proximity protocol coexist is considered in~\cite{ref:rel10}. They proposed a greedy algorithm of admission control of LTE proximity services to maximize the reduction of latency caused by two proximity protocols. In~\cite{ref:rel10-1}, a computation offloading scheme for mobile edge computing technology with vehicle devices. The authors of~\cite{ref:rel11} proposed a fast discovery and radio resource allocation algorithm to minimize the latency of proximity communications. Deep reinforcement learning is used to allocate radio resource and transmission power of D2D nodes in~\cite{ref:rel12}. The latency can vary depending on where the measurements are placed in the communication procedure. 
	    
	\subsection{Resource allocation with deep learning}
	  
    Since deep learning has produced innovative results in the computer visions~\cite{ref:dl1}, many researchers have studied the application of deep learning to wireless communications. Currently, results using deep learning in each field of wireless communication are being announced. In resource allocation of wireless communications, there are also several impressive results. In the first generation, resource allocation and power control based on deep learning are studied with simple problem and labels from a known algorithm. The authors of~\cite{ref:wdl1} proposed a power control scheme using DNN. They conduct WMMSE~\cite{ref:wmmse} to get labels, then train DNN to predict the labels with all channel information. It is helpful to reduce computation time.
    Next studies had been conducted for more complex problems. In~\cite{ref:wdl2}, the authors use Convolutional neural networks (CNN) to inference the labels with incomplete channel information. The authors of~\cite{ref:wdl3} use Recurrent neural networks (RNN) to solve Non-orthogonal multiple access (NOMA) problem.

    For the cases of researches about D2D related, the authors in~\cite{ref:intlink} used deep learning for intelligent link adaption to determine transmission rate. A V2V resource allocation is proposed with deep Q-networks (DQN) in~\cite{ref:rel12}. It adopts a way that one of several given options is chosen because DQN is a discrete decision algorithm. However, the transmit power is a continuous variable. Thus, there is room for further performance improvement. In this paper, we suggest a transmit power allocation scheme that is available with continuous action spaces. Meanwhile, the authors in~\cite{ref:wdl4} proposed a D2D resource allocation with deep learning. They do not use labels but optimize the objective function directly using deep learning. The different from our proposed scheme is that it is based on central manner. For the IoT-D2D environments, a distributed scheme has to be seriously considered.  

	\section{Proposed scheme}
	
	     In this section, we describe the proposed Distributed Power Allocation method using DNN with Interference to eNB Constraint (DPADIC). 
	
	\subsection{System model}
	It is assumed that orthogonal frequency division multiplexing access (OFDMA) is used in the considering cellular networks. It has $N$ orthogonal subcarriers, which are non-overlapped. The spanned bandwidth is smaller than the channel coherence bandwidth, so the spectrum is regarded as flat. We consider a set $\mathcal{N} = \{1,...,N\}$ of shared OFDMA channels, as well as a set of D2D device pairs, $\mathcal{K} = \{1,...,K\}$. The pair of D2D devices consists of a transmitter and receiver, which are considered to be in perfect synchronization. Likewise, we consider multi-cell environments with $B$ cells. The set of eNB is $\mathcal{B} = \{1,...,B\}$. As shown in~\cite{ref:d2d_conv}, a received signal $Y_{n,k,k}$ on link $n$ can be expressed as follows:
		
	\begin{equation}
	Y_{n,k,k}= H_{n,k,k} S_{n,k,k} + \sum_{i \in \mathcal{K}, i \neq k} H_{n,i,k}S_{n,i,k} + W_{n,k,k}
	\end{equation}
	
	\noindent where $H_{n,k,k}$ means the complex channel gain between the transmitter and receiver of D2D device pair $k$. The $H_{n,i,k}$ is also the complex channel gain from the transmitter of D2D pair $i$ to the receiver of D2D pair $k$. $S_{n,k,k}$ is the symbol of transmission. $W_{n,k,k}$ is an additive noise from zero-mean Gaussian distribution with variance $(\sigma_{n,k})^2$. Therefore, the spectral efficiency $T_k$ at a receiver of D2D pair $k$ is expressed as follows:
	
	\begin{equation}
	T_k(\mathbf{p}_k) = \sum_ {n\in \mathcal{N}} \log_2 (1+ \frac{(H_{n,k,k})^2 p_{n,k}}{\sum_{i \in \mathcal{K}, i \neq k } (H_{n,i,k})^2 p_{n,i} + (\sigma_{n,k})^2 })
	\label{eq:throughput}
	\end{equation}
	
	\noindent where $p_{n,k}$ is transmit power for D2D pair $k$ on channel $n$. $\mathbf{p}_k$ is a set of $p_{n,k}$ on each channel, $\mathbf{p}_k = \{p_{1,k}, p_{2,k},...,p_{N,k}\}$. The proposed scheme aims to maximize the sum of D2D throughput while maintaining the following two constraints: power constraint, and interference to eNB constraint. Therefore, the objective function and constraints can be derived as:
	
	
	\begin{subequations}
	\label{eq:maxi}
	\begin{align}
	&\max \quad {\sum_{k \in \mathcal{K}} T_k(\mathbf{p}_k)}&\\
	    &\text{subject to} \nonumber &  \\
	    &\quad  \sum_{n \in \mathcal{N}} p^n_k \leq P_{max}, \quad  k \in \mathcal{K} & \\
	    &\quad  \sum_{k \in \mathcal{K}} (H_{n,k,c})^2 p^n_k, \quad  n\in \mathcal{N} & 
	\end{align}
	\end{subequations}

	\noindent where $P_{max}$ is the power limitation of each D2D transmitter, and $Q_{max}$ is the interference to eNB constraint per channel. The maximum power constraint means that the total transmit power per user cannot exceed a given limit $P_{max}$. Also, the interference constraint means that the interference experienced at the eNB cannot exceed the threshold $Q_{max}$. 
	
	\subsection{Deep learning model for distribute power allocation with interference constraints}
	After the training phase in a central machine, all D2D devices have the same deep learning model. The model can autonomously infer transmit power with the location information of a transmitter and a receiver only. The proposed distributed decision scheme can maximize the total D2D rate in the multi-cell environment while maintaining interference constraints. 
	
	The deep learning model uses a pair of location information to derive a pair of transmit power. In the training phase, the inferred transmit powers from every device are collected to calculate the sum of throughput. The sum of throughput is used to update the deep learning model. In conclusion, the model is trained taking into consideration the inferred power on the data link and the interference on other data links. After the training phase in a central machine, all D2D devices have the same deep learning model. The model can autonomously infer transmit power with the location information of a transmitter and a receiver only. The proposed distributed decision scheme can maximize the total D2D rate in the multi-cell environment while maintaining interference constraints.

	

	

	\Figure[h]()[width = 0.99\columnwidth]{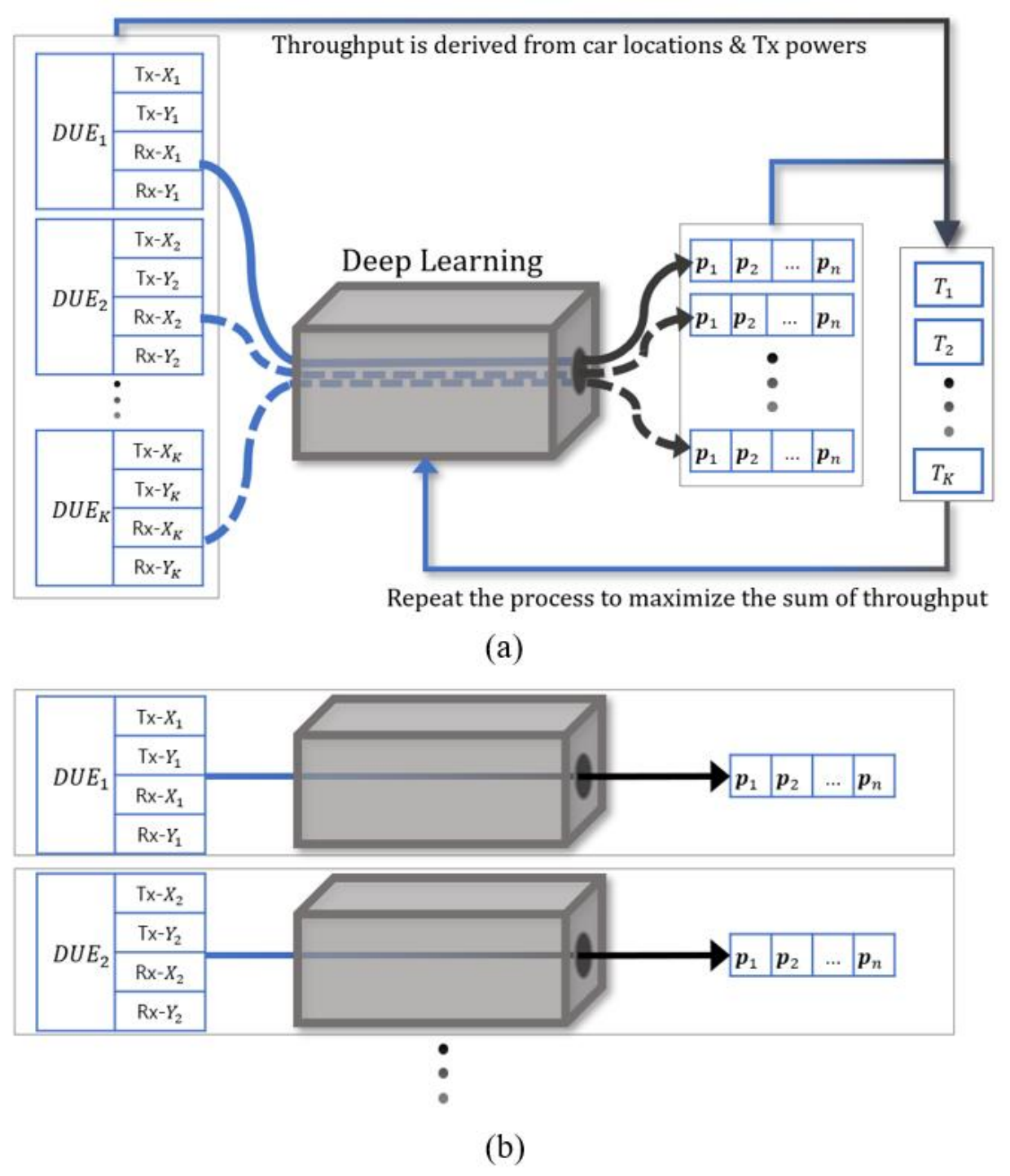}{Distributed learning model which autonomously determines transmit power to maximizes the total throughput. (a): Training phase, (b): Inference phase. \label{fig:main}}



		
	\subsubsection{Distributed deep learning architecture}
		Fig.~\ref{fig:main} shows that the distributed deep learning architecture. There are two phases: training phase in (a) and inference phase in (b). In the training phase, a deep learning model is trained with all location information of whole D2D devices in cells. For example, a D2D pair $DUE_1$ has four number: (x,y) of transmitter and receiver. The four numbers are a unit of data. The K units of data are used to train the deep learning model as independent input data. It means that the model infers transmit power differently to each pair. After that, the inferred transmit power are evaluated with the sum of throughput and constraints. The throughput of each pair is not maximized independently. The deep learning model is trained to maximize the sum of the throughput. After the training phase in a single machine, all D2D devices have the same deep learning model. Consequently, the models autonomously determine the transmission power of each D2D device only with local location information while maximize global objective function: the sum rate of D2D in multi-cell. In~\cite{ref:T-DNN}, the similar concept has been introduced but the proposed scheme has advanced features. The biggest difference is that we use one model. It simplifies overall training process and enhances feasibility of the proposed scheme. If multiple models are adopted for difference devices, then each model is trained by different data set. In that case, it is ambiguous that which model should be given to which device. If online learning is adopted instead of pre-trained model, another problem can be occurred. In online learning, deep learning can be affected by too much initiative data. Overfitting can also be occurred in the initiative data. If the multiple models use the same data set during training phase, those would become the same model consequently. Thus, one large model is more efficient to achieve the same result compared to cooperative multiple models. The distributed architecture is described as follows. Typically, $\theta$ is defined as a policy parameter. The policy for a D2D pair $k$ is $\theta_k$. Then, the optimal set $\bar{\mathbf{\theta}}^*_k$ can be defined as
	
	\begin{equation}
	\bar{\mathbf{\theta}}^*_k = \operatorname*{argmax}_{\bar{\mathbf{\theta}}_k} \sum_{k \in \mathcal{K}, \theta_k \in \bar{\mathbf{\theta}}_k} T_k (\mathbf{p}_k (\theta_k) )
	\label{eq:optimal_theta}
	\end{equation}
	
	\noindent where $\mathbf{p}_k (\theta_k)$ is transmit power which is derived from the policy $\theta_k$ for D2D pair $k$. Each element of $\bar{\mathbf{\theta}^*_k}$ are different from each other to optimize Eq.~\ref{eq:maxi}. However, the proposed model pursues that every D2D device has the same machine to determine their transmit powers to achieve the near optimal spectral efficiency. It means that every device in the same set $\mathcal{K}$ has the same $\theta_{\mathcal{K}}$ as    
	
	\begin{equation}
	{\mathbf{\theta}}^*_{\mathcal{K}} = \operatorname*{argmax}_{\theta} \sum_{k \in \mathcal{K}} T_k (\mathbf{p}_k (\theta) )
	\end{equation}
	
	\noindent where ${\mathbf{\theta}}^*_{\mathcal{K}}$ is the optimal $\theta_{\mathcal{K}}$. Note that all pairs of devices $k$ have the same ${\mathbf{\theta}}^*_{\mathcal{K}}$ in $\mathcal{K}$. Also, the results of ${\mathbf{\theta}}^*_{\mathcal{K}}$ should approximate the result of the optimal set $\bar{\theta}^*_k$ as
	
	\begin{equation}
	\sum_{k \in \mathcal{K}} T_k (\mathbf{p}_k (\mathbf{\theta}^*_{\mathcal{K}}) ) \lesssim \sum_{k \in \mathcal{K}, \theta^*_k \in \bar{\mathbf{\theta}}^*_k} T_k (\mathbf{p}_k (\theta^*_k) )	\label{eq:near1}
	\end{equation}
		
	\noindent Extensively, a set of $\mathcal{K}$ can be defined as $\bar{\mathcal{K}} = \{ \mathcal{K}_1, \mathcal{K}_2, ..., \mathcal{K}_\mathbf{B} \}$ where $\mathbf{B}$ is the number of sets. From that, the $\theta$ also can be redefined for $\bar{\mathcal{K}}$ as
			
	\begin{equation}
	{\mathbf{\theta}}^*_{\bar{\mathcal{K}}} = \operatorname*{argmax}_{\theta} \sum_{\mathcal{K} \in \bar{\mathcal{K}}} \sum_{k \in \mathcal{K}} T_k (\mathbf{p}_k (\theta) )
	\end{equation}

	\noindent Finally, we define the target $\theta$ which is independent to distributions of other devices while satisfy
		
	\begin{equation}
	\sum_{\mathcal{K} \in \bar{\mathcal{K}}}\sum_{k \in \mathcal{K}} T_k (\mathbf{p}_k (\mathbf{\theta}^*_{\bar{\mathcal{K}}}) )  \lesssim \sum_{\mathcal{K} \in \bar{\mathcal{K}}}\sum_{k \in \mathcal{K}, \theta^*_k \in \bar{\mathbf{\theta}}^*_k} T_k (\mathbf{p}_k (\mathbf{\theta}^*_{\mathcal{K}}) )
	\end{equation}
	
	\noindent It is difficult to approximate  $\theta^*_{\bar{\mathcal{K}}}$ to have the result of the optimal $\bar{\theta}^*_k$ in Eq.~\ref{eq:optimal_theta}. It is why deep learning should be adopted. Therefore, the policy $\theta$ can be redefined as a set of weights and bias in the DNN, $ \{\textbf{W}, \textbf{b} \}$. According to $\theta$, the neural network can determine the transmit power $\mathbf{p}$ so the $\theta$ is still the policy parameter. Thus, the $\mathbf{p}$ can be redefined with DNN as
	
	\begin{equation}
	\mathbf{p}_k ( \theta^*_{\bar{\mathcal{K}}}) = DNN(k, \theta^*_{\bar{\mathcal{K}}} )	
	\end{equation}
	
	\noindent where DNN is a neural network, which can determine the transmit power $\mathbf{p}_k$ based on the D2D pair $k$ and the weights and bias set for $\bar{\mathcal{K}}$. Deep learning is a process for finding the optimal $\theta$. Intuitively, if $ \theta^*_{\bar{\mathcal{K}}}$ is sufficiently large, it can include all the meanings of the elements of  $\bar{\theta}^*_k$.
	
%
	\subsubsection{Cost function}
	In DPADIC, two constraints should be reflected to the cost function: i) transmitting power constraints, ii) interferences to eNB constraints. We adopt the Lagrange function to express the two constraints in the cost function. In deep learning process, a cost function defines a way to give benefit or penalty to update DNN. In other words, a cost function can be customized if it can give benefit or penalty. Therefore, we use throughput directly to the cost function of deep learning itself in the proposed scheme, as shown in Eq.~\ref{eq:throughput}. Thus, labels of data are not required. Although throughput and constraints are non-convex, it can be approximated by using deep learning. The power constraint $\eta_p$ is expressed as follow:
	
	\begin{equation}
	\eta_p(\mathbf{p}_k) =\sum_{k \in \mathcal{K}} \log_2(1+\frac{ReLU(\sum_{n \in \mathcal{N}} p_{n,k} - P_{max})}{P_{max}})
	\label{eq:power_const}
	\end{equation}
	
	\noindent where ReLU is the rectified linear unit (ReLU) function which is $ReLU(x)=max(0,x)$. If the sum of the transmit power of a D2D transmitter is under the threshold $P_{max}$, $\eta_p$ would be $0$. Therefore, it only delivers a penalty if the transmit power of the transmitter exceeds the constraint. Besides, it is designed like Shannon capacity for being easy to make similar scale. Note that $ReLU(\sum_{n \in \mathcal{N}} p_{n,k} - P_{max})$ is a ratio unit as similar to the definition of SINR. If the difference of scale is too large between independent terms in a cost function, deep learning cannot maintain balances of terms while training. Traditionally, additional constants, e.g) Lagrange multipliers, are used to balance the terms. We also adopt them but finding appropriate multipliers for deep learning is another challenge. Instead of that, we make constraints having similar scales to Shannon capacity. There are two points: using ReLU and similar form to Shannon capacity to make easy to find appropriate Lagrange multipliers.

	The interference to eNB constraint is also designed in a similar way like that to the power constraint. Before defining the constraint formula, the term of interference to eNB should be defined, which can be expressed as follows:
	\begin{equation}
	Q_{n,k,b}(\mathbf{p}_k) =\sum_{k \in \mathcal{K}}(H_{n,k,b})^2p_{n,k}
	\label{eq:if_const}
	\end{equation}
	\noindent where $b$ means an eNB, and it is $b \in \mathcal{B}$. According to Eq.~\ref{eq:maxi}, the interferences to eNB constraints are set for each channel. Note that the noise is not adopted for the formula. This formula aims to estimate the impact of each D2D transmitter on the eNB. Thus, the random noise factor should be ignored. Therefore, the interference to eNB constraints, $\lambda_{if}$, can be formulated as follows:
	
	
	\begin{equation}
	\eta_{if}(\mathbf{p}_k) = \sum_{k \in \mathcal{K}} \sum_{b \in \mathcal{B}} \sum_{n \in \mathcal{N}} \log_2(1+\frac{ReLU(Q_{n,k,b}(\mathbf{p}_k) - Q_{max})}{Q_{max}})
	\end{equation}
	Finally, the cost function, $C$, of the proposed method can be described as follows:
	
	\begin{equation}
	\label{eq:cost}
	C(\mathbf{p}_k) = -\sum_{k \in \mathcal{K}} T_k(\mathbf{p_k}) + \lambda_{if} \eta_{if}(\mathbf{p}_k)  +  \lambda_{p} \eta_{p}(\mathbf{p}_k)
	\end{equation}
	
	\noindent where $\lambda_{if}$ and $\lambda_{p}$ are Lagrange multipliers. Finding appropriate $\lambda_{if}$ and $\lambda_{p}$ are easy because they have a similar form to the objectives and ReLU in $C$.

	
	\subsection{Deep learning process}
	We adopt a multi-layered neural networks (MLP) to predict transmit powers. The number of features in an input data are only four, which are the locations of transmitter and receiver, so other extended deep learning architectures such as Convolutional neural network (CNN) do not need to be considered. For activation function, we use a sigmoid, which is $\frac{1}{e^x+1}$, instead of the ReLU. The defined problem is a regression problem. Thus, ReLU, which is a concept that identifies the required partial feature, is not appropriate. Sigmoid is suitable for the proposed method because it can deliver gradient to the previous layer with a back-propagation algorithm while preventing divergence of the neural network. If a vanishing problem is revealed, ResNet~\cite{ref:resnet} can be used to deal with it but such a complicate network is not required because the input data consists of four features. In particular, the proposed method is more sensitive to germination, as there is a constraint for maximum power. 
	
	The learning process in the proposed scheme is similar to typical deep learning, except that simulation can be included in the training phase. In the proposed scheme, the deep learning process is merged with the simulation, which generates the location information of D2D nodes as input data to the learning process. It is a distinguished feature of the proposed scheme compared to typical deep learning process. 
	
	Input data and labels are important components for successful deep learning. Deep learning is trained to deliver output data to be similar with the labels of the input data. Thus, a successful learning process may not be guaranteed for the input data without labels. The problem to be solved in this paper corresponds to this case. The system cannot know the proved optimal solution before the resource and power allocation.

	Instead of labels from the proved optimal solution, we use the objective function Eq.~\ref{eq:cost} as the cost function of deep learning. Because of this, the simulation generates new data every time for training batch data. Thus, the simulation generates as much input data as required at each iteration. It means that there is no overfitting. The detailed learning process is described in Algorithm~\ref{alg:alg1}.


	We adopt Xavier initiation~\cite{ref:xavier}. $n\_epoch$ is the number of iterations. The simulation is designed to deliver a batch, which is a set of input data. The size of a batch is given as batch\_size. Train() function is the actual training part in~\ref{alg:alg1}. Get\_Throughput(X,P) delivers the throughput as defined in Eq.~\ref{eq:throughput}. Finally, the throughput results are included in a set $Throughput$. The throughput results are collected in order of iteration in the set $Throughput$. Train() inferences the power set $P$ with input data $X$ and $\theta$. Then, the cost function is defined as $c$ with the input data as $X$ and the predicted power as $P$. The cost function is the main part of this train function. It is implemented using Eq.~\ref{eq:cost}.
	
	\begin{algorithm}[tb]
		\scriptsize
		\SetAlgoLined
		\SetKwInOut{Input}{Input}\SetKwInOut{Output}{Output}%
		\caption{Proposed scheme}
		\label{alg:alg1}
		\Input{input\_size=4, output\_size=8, width, depth, n\_epoch, batch\_size}
		\Output{Throughputs}
		\nl $\theta$ = Xavier\_initiation($\theta$) \\ 
		\nl \For{$i$ = 1, ... , n\_epoch}{
			\nl 	$X$ = Simulation(batch\_size) \\
			\nl 	$P$, $\theta$ = Train($X$, width, depth, batch\_size, $\theta$) \\
			\nl 	$T$ = Get\_Throughput($X$, P) \\
			\nl 	Throughputs.append($T$) 
		}
		\nl \KwRet{Throughputs}
		\setcounter{AlgoLine}{0}
        \\\hrulefill
        \\
		\SetKwFunction{al}{Train}{}{}
		\SetKwProg{malg}{Function}{}{}
		\SetKw{mstate}{\malg{\al{}}{}{}}
		\malg{\al{}}{}{} 
		\SetKwInOut{Input}{Input}\SetKwInOut{Output}{Output}%
		\Input{$X$, batch\_size, width, depth, $\theta$}
		\Output{$P$}
		\nl $P$ = Inference(X, width, depth, batch\_size, $\theta$) \\
		\nl $c$ = cost(X, P) \\
		\nl $\theta$ := AdamOptimizer(c)~\cite{ref:adam}\\
		\nl \KwRet{P, $\theta$}
		\setcounter{AlgoLine}{0}
		\SetKwFunction{al}{Inference}{}{}
		\SetKwProg{malg}{Function}{}{}
		\malg{\al{}}{}{} 
		\Input{$X$, width, depth, batch\_size, $\theta$}
		\Output{$Y_{pred}$}
		\nl $W$, $S$, $Z$ = $\theta$ \\
		\nl $X$ = reshape(X, [batch\_size * K, 4])\\
		\nl \For{j = 0, ..., depth-1}{
			\nl 	$X = dense\_layer(X, width)$ \\
			\nl 	$X = batch\_norm(X)$~\cite{ref:batch_norm}\\
			\nl 	$X = \frac{1}{1 + e^X}$ 
		} 
		\nl $X = dense\_layer(X, output_size)$ \\
		\nl $X = \frac{1}{1 + e^X}$ \\
		\nl $P$ = reshape(X, [batch\_size, K, 8]) \\
		\nl \For{p = each element of $P$}{
			\nl 	$p$ = $p \times 170 - 150$ 
		}
		\nl \KwRet{$P$}\\
	\end{algorithm}
	
			\begin{table}[tb]
			\centering
			\caption{Parameters of simulation}
			\begin{tabular}{|c | l | c|}
				\hline
				\textbf{Term} & \textbf{Description} & \textbf{Value} \\
				\hline
				\hline
				$R$ & cell radius & 500m \\
				\hline
				$D_{max}$ & maximum distance between D2D pair & 100m \\
				\hline
				$B$ & the number of cells & 3, 7 \\
				\hline
				$N$ & the number of subchannels & 8 \\
				\hline
				$K$ & the number of D2D pairs in a cell & 8 \\
				\hline
				$P_{max}$ & maximum transmit power & 0.25 W\\
	            \hline
	            $\alpha$ & path loss exponent & 4 \\
	            \hline
	            $\sigma$ & shadowing standard deviation & 8 dB \\
	            \hline
			\end{tabular}
			\label{tab:parm_sim}
		\end{table}
		
	\Figure[!t]()[width=1.0\textwidth]{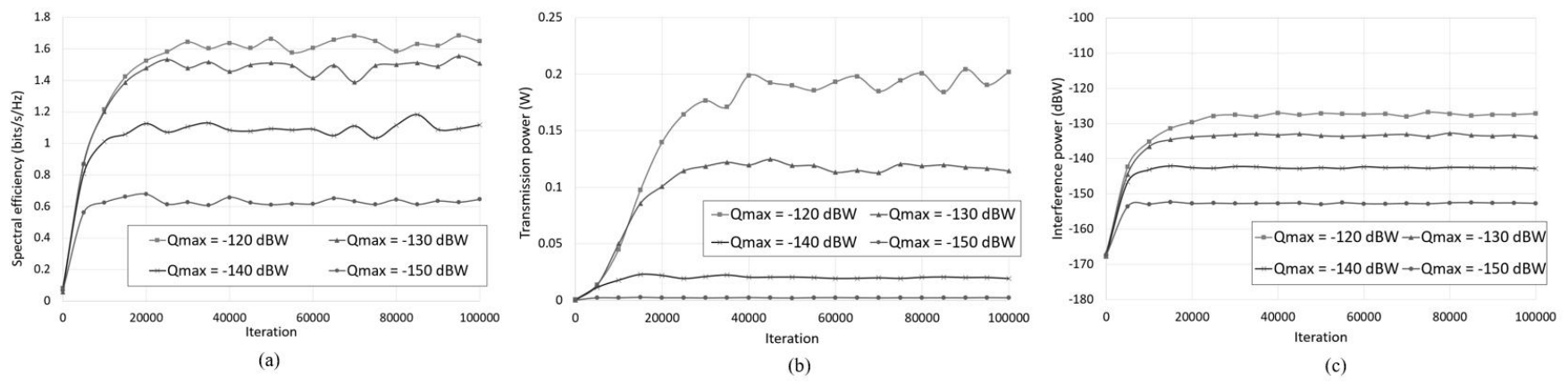}{(a): Spectral efficiency, (b): Transmit power of each D2D transmitter and (c): Interference experienced at the eNB where $B=3$. \label{fig:b3}}
	
	\Figure[!t]()[width=1.0\textwidth]{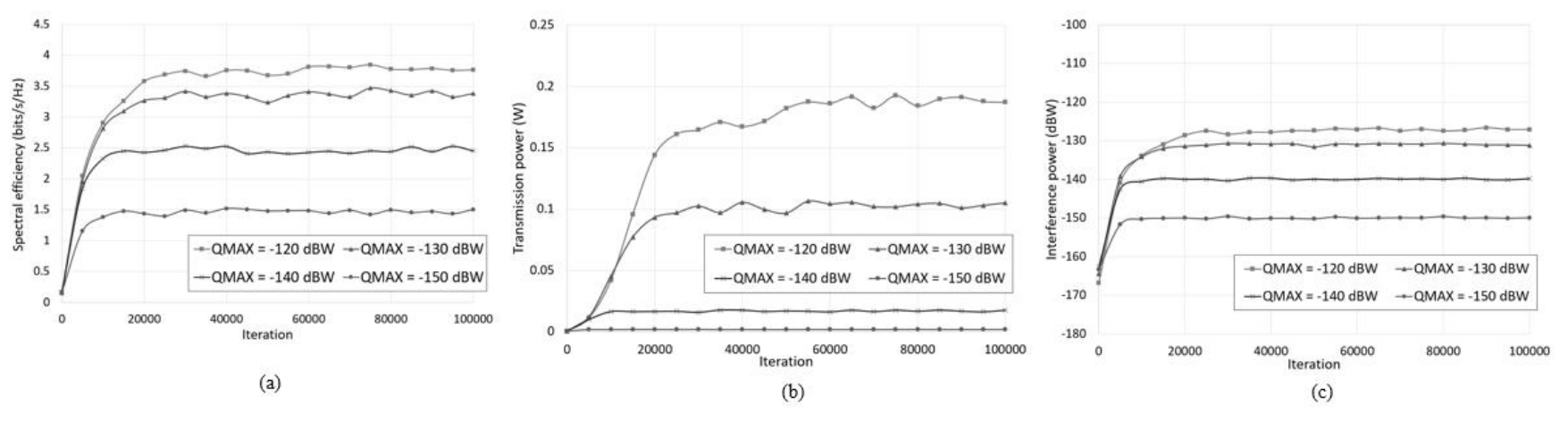}{(a): Spectral efficiency, (b): Transmit power of each D2D transmitter and (c): Interference experienced at the eNB where $B=7$. \label{fig:b7}}

	$X$ and $P$ may have several data sets because the several input data sets are trained simultaneously. In the cost function, Eq.~\ref{eq:cost} of each input data set is derived 
	, and the results are averaged. We also use the Adam optimizer in~~\cite{ref:adam} to adjust $\theta$, which deals with the cost function itself, not the result of the cost function. The Adam optimizer differentiates the cost function to trace the changes. Consequently, $\theta$ is gradually changed by the optimizer to minimize the cost function. In Inference(), the reshape function is used to change the shape of the input data.

	The first shape of the input data is [batch\_size, K, 4], which means that there is a number of batch\_size and an input data set has K number of D2D pairs. A D2D pair has four features: x, y of the transmitter and receiver, respectively. It should be changed to [batch\_size $\times$ K, 4] because each D2D pair data should be independent of distributed learning. Thus, there are $batch\_size \times K$ D2D pairs. The function dense\_layer() is a neural layer. After the output layer, it should be rescaled between -150 and 20, because the unit of output power is in dBm. 
	
According to the proposed scheme, it can reflect large-scale fading including path-loss and shadowing. The path-loss can be modeled as a function of distance statistically. Because distance can be easily implied from the location, we can understand that the computation of path-loss is implicated inside of the neural network.

The shadowing effect is dependent on the location of a device because it is closely related to physical obstacles to signal, such as buildings or trees. In simulations, however, it is difficult to reflect the effects of random variables based on a neural network if the random value is not one of the input data. This problem can be mitigated to use enough practical data in the learning process or adopt more detailed channel model.

Small scale fading is usually defined with a normal distribution, and thus it is impossible to estimate small fading effect with only location information. However, the small scale fading can be negligible because of the purpose of the proposed scheme: drastically shortening the resource allocation latency instead of focusing on a near-optimal solution of the non-convex problem. Thus, to consider the small scale fading is out of scope in this paper but we remain it as a future work. The problem of adopting small fading to D2D communications can be covered by applying recent works to estimate the channel models~\cite{ref:v2v_modeling}.


	\section{Results}
	We consider the same experimental assumption with~\cite{ref:d2d_conv}. The simulation parameters are summarized in Table~\ref{tab:parm_sim}. We assume hexagonal cells with radius R = 500m. The maximum distance between D2D pairs is $D_{max} = 100 m$, while they are uniformly distributed in $[0, D_{max}]$. In addition, we consider multi-cell cases: $B = 3$ and $B = 7$ where $B$ is the number of cell. The number of D2D pairs is 8 per cell. Thus, the number of D2D pairs $K$ is $8 \cdot B$. The number of OFDMA subchannels $N$ is set to 8, and then the spectral efficiency $\eta$ is derived as $\eta = \frac{\sum T_k}{(K \times N)}$. The maximum transmit power constraint $P_{max}$ is set to 0.25 W. The channel attenuation is expressed by the path loss with distance, including shadowing and fading. The path loss exponent $\alpha$ is 4, with shadowing and standard deviation $\sigma = 8 dB$ on log normal distribution. The additive zero-mean Gaussian noise in the cellular network from D2D is set to -130 dBW in~\cite{ref:winner}. This simulation is implemented using Tensorflow~\cite{ref:tensorflow}.

	\begin{table}[tb]
			\centering
			\caption{Parameters of DNN}
			\begin{tabular}{c c}
				\hline
				Width & 1500 \\
				Depth & 7\\
				Batch size & 50 \\
				The number of iterations & 100K \\
				Learning rate & 0.0001 \\
				\hline
			\end{tabular}
			\label{tab:parm_best}
		\end{table}

	\Figure[h]()[width = 0.99\columnwidth]{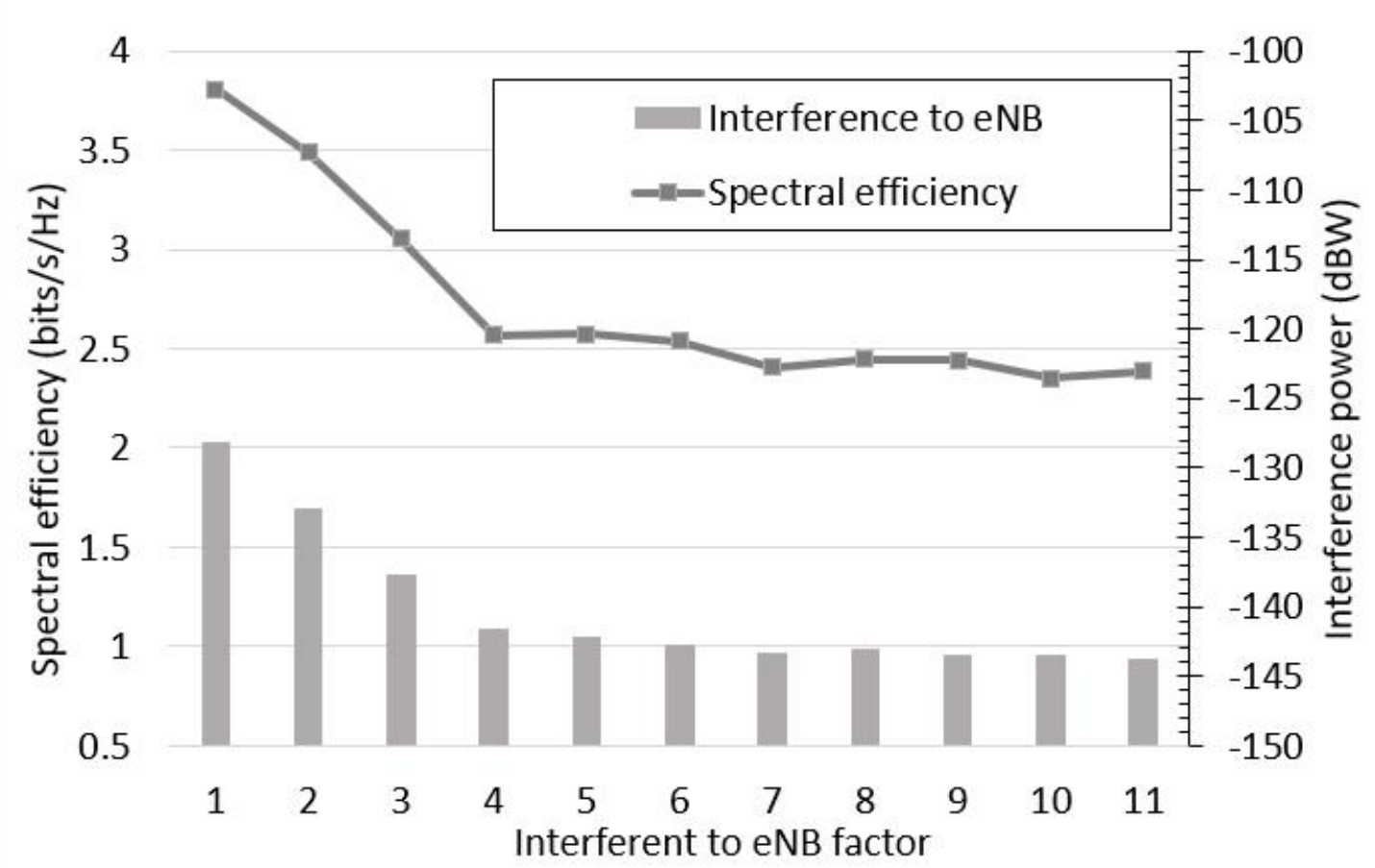}{Spectral efficiency with interference to the eNB constraint factor, where $B = 3$ and QMAX = -140 dBW. \label{fig:140ef}}

	\Figure[h]()[width=0.99\columnwidth]{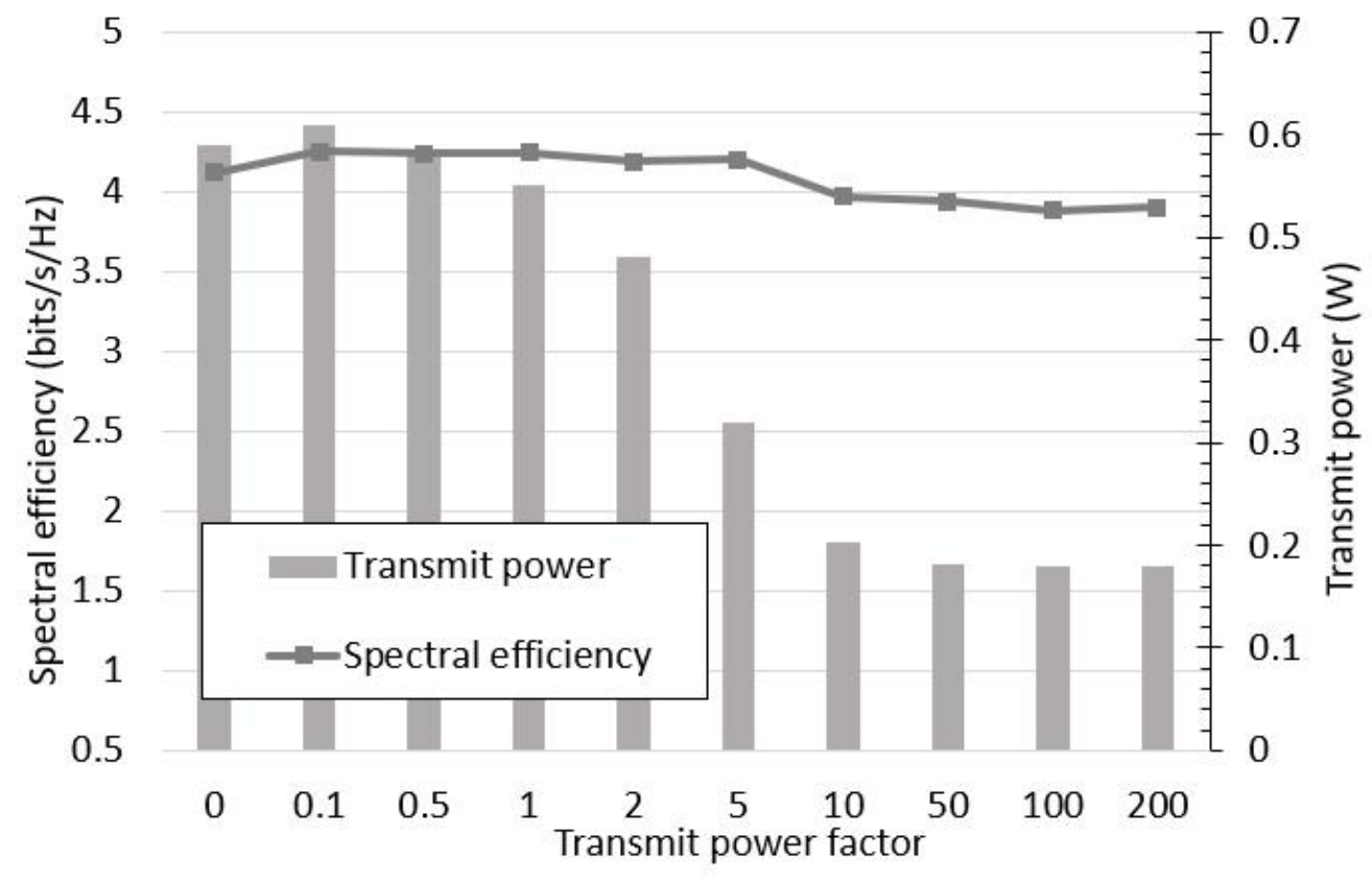}{Spectral efficiency with the transmit power factor, where $B = 3$ and QMAX = -110 dBW. \label{fig:110p}}

	We use 50 data sets for a batch and total iterations are 100K. Thus, we use 5M cases of drops for training and there are no duplicated data because the data sets are newly generated in every iteration. The learning rate of the optimizer is 0.0001. If the learning rate is increased, DNN can attain a converged D2D rate earlier with fewer iterations. However, the final converged D2D rate may be decreased. Hyper parameters are 7 layers and 1500 perceptrons per layer. The size of neural network can be regarded as too large, but it is not a problem with computing power with this entry-level GPU. With these parameters, the learning time is about 3~4 hours. We use I7-6700K processors and a GTX 1080 Ti. It is another area of deep learning research that producing the same result with a smaller neural network. In addition, the function of inference is able with CPU, which means that it requires less computing power. Those deep learning parameters are summarized in~\ref{tab:parm_best}.

	Fig.~\ref{fig:b3} and Fig.~\ref{fig:b7} describe the performance of the proposed scheme where $B = 3$ and $B = 7$ respectively. They tend to converge to a constant value after 30K iterations. Smaller $Q_{max}$ cases tend to be converged earlier because the initial transmit power is close to zero, as shown in Figs.\ref{fig:b3}-(b)and~\ref{fig:b7}-(b). We set the range of power between -150 and 20 dBm. The initial powers are set near the middle of the range. The power is increased to find a better throughput by using the optimizer. Figs.~\ref{fig:b3}-(c) and~\ref{fig:b7}-(c) shows that DNN obtains the converged throughput while maintaining the constraint of interference to eNB.

	In the proposed scheme, there are two significant parameters for adopting constraints, $\lambda_{if}$ and $\lambda_p$. They should be determined manually, but it is not difficult because the valid range of the parameters is wide enough. Fig.~\ref{fig:140ef} show the effects of the interference to the eNB constraint factor, $\lambda_{if}$. If too small $\lambda_{if}$ is used, the interference to eNB constraints can be ignored. In that case, it is more profitable to ignore $\lambda_{if}\eta_{if}$ in minimizing the cost, though DNN takes the penalty from $\lambda_{if}\eta_{if}$. Thus, the spectral efficiency $T$ is high but it is not valid because the interference to eNB exceeds the limit, $Q_{max}$. If $\lambda_{if}$ is high enough, DNN cannot ignore the constraint. Then, DNN should maintain the constraints with reduced transmit power. If a much higher $\lambda_{if}$ is used, $T$ can be reduced, but the falling is not meaningful. Note that $\eta_{if}$ includes ReLU function. It turns off the constraint if it does not exceed the threshold. Because of this, an effect of a high $\lambda_{if}$ is limited. However, D2D transmitters are dropped randomly, and it may be very close to the eNB. Thus, there can be a few cases of exceeding $Q_{max}$ though it has a very small transmit power. The cases affect the results. Consequently, $T$ can be reduced slightly with larger $\lambda_{if}$.

	\begin{figure}[tb]
	\centering
	\includegraphics[width=0.9\columnwidth]{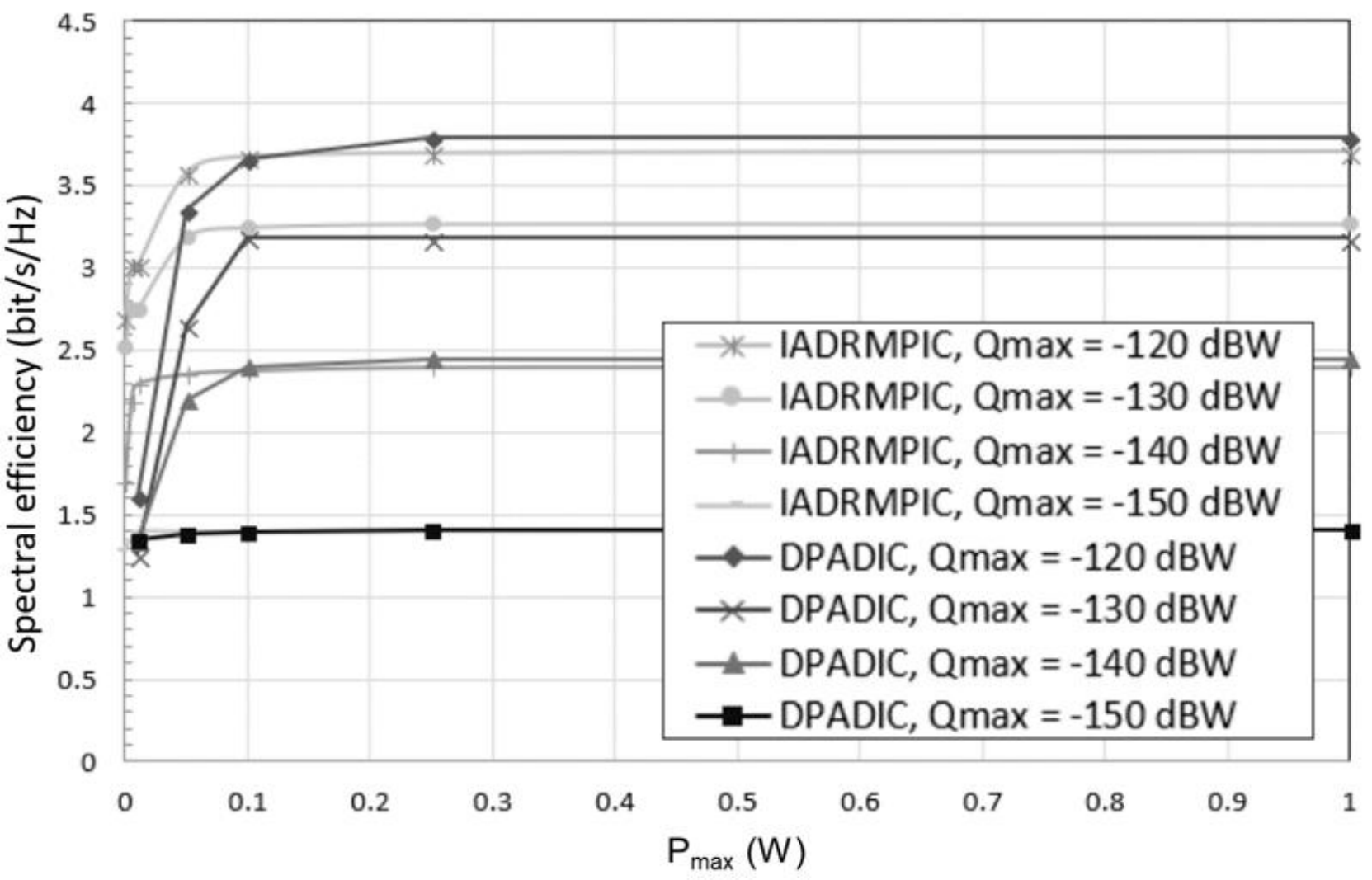}
	\caption{Comparison to IADRMPIC~\cite{ref:d2d_conv} with various $P_{max}$, where $B = 7$.}
	\label{fig:pw_compare}
	\end{figure}
	
	\begin{figure}[tb]
		\centering
		\includegraphics[width=0.9\columnwidth]{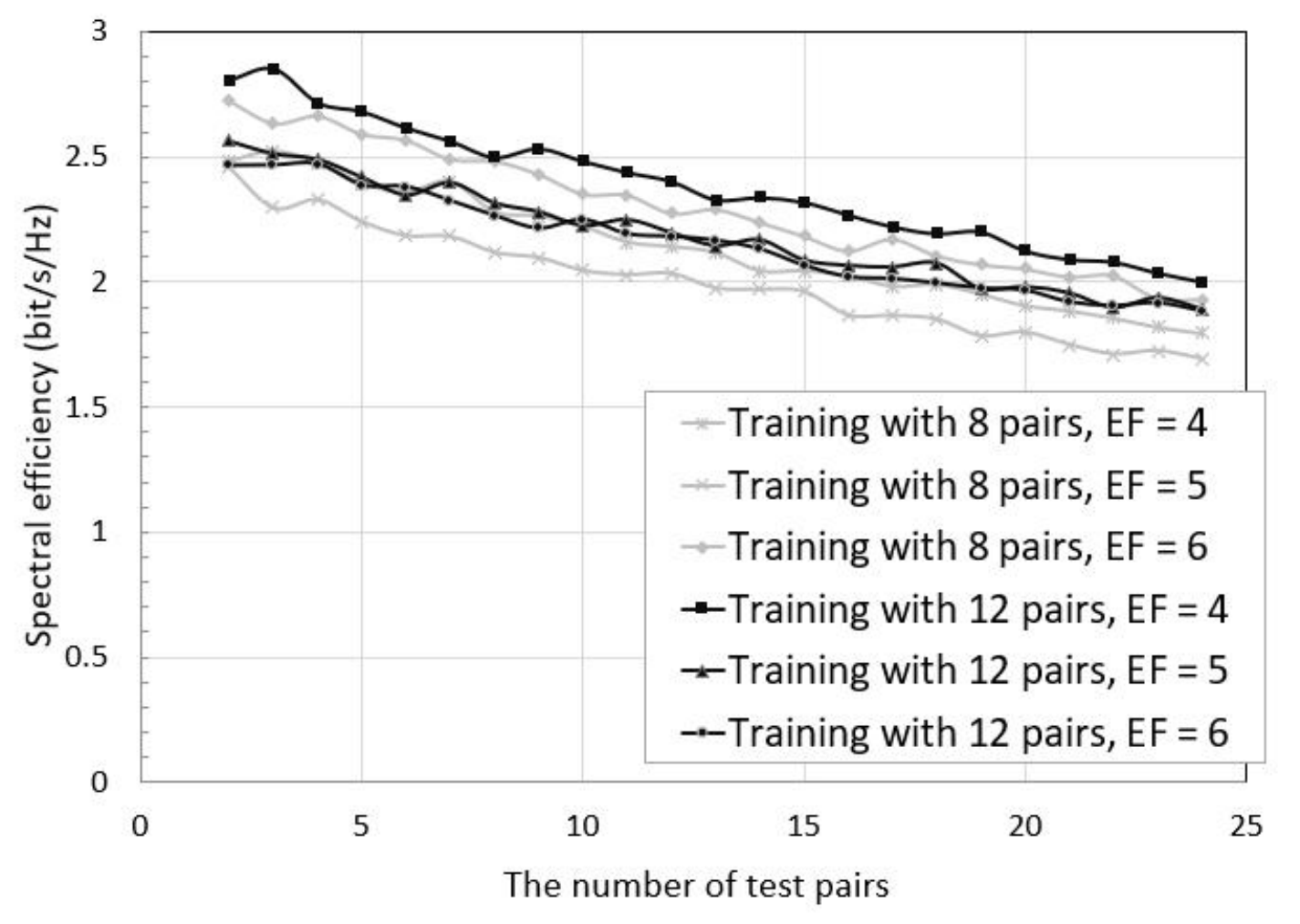}
		\caption{Spectral efficiency with various number of devices, where $B = 3$.}
		\label{fig:scala}
	\end{figure}
	
	\begin{figure}[tb]
		\centering
		\includegraphics[width=0.9\columnwidth]{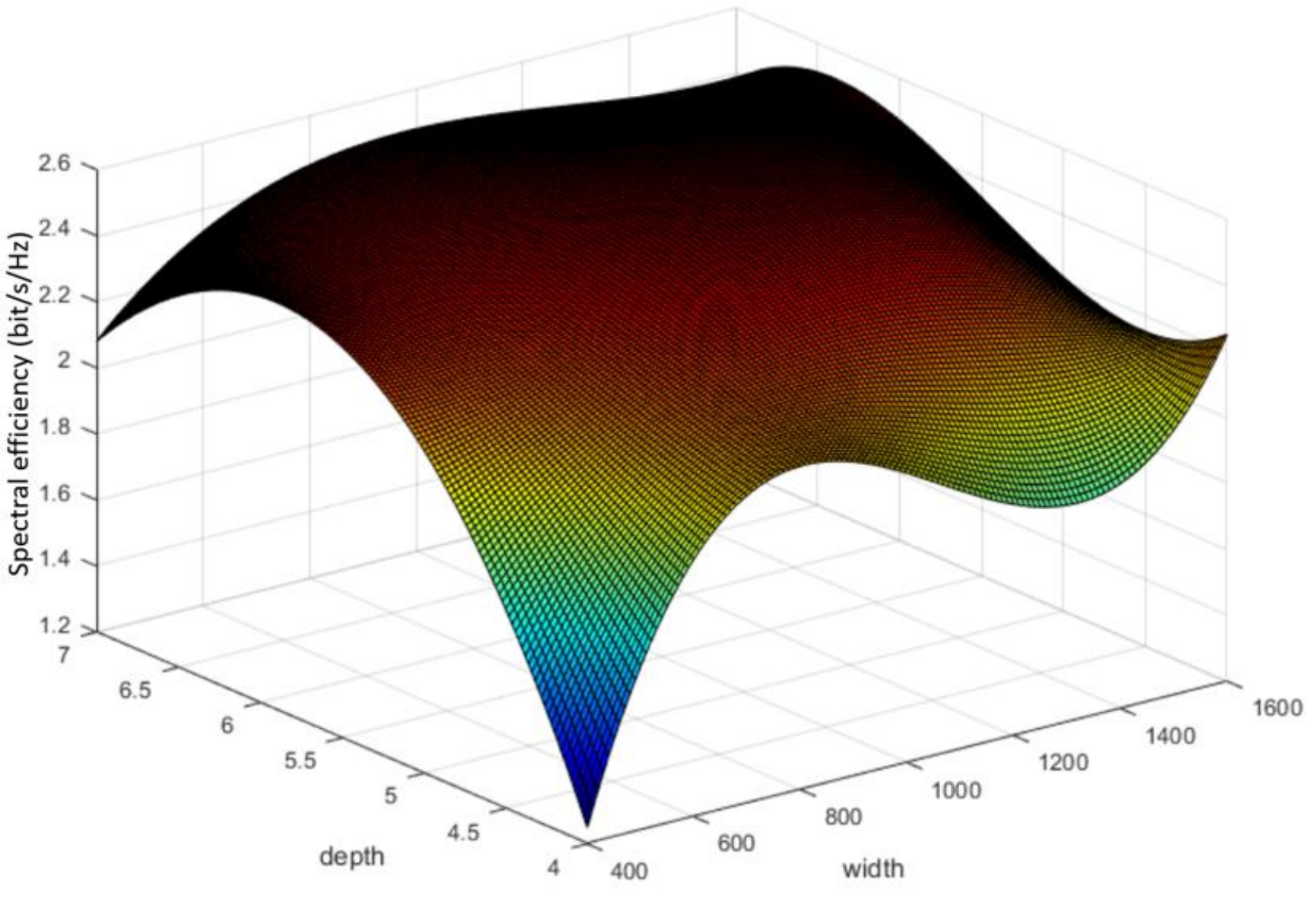}
		\caption{Spectral efficiency with various hyper parameters, where $B = 3$ and the number of devices per a cell = 16.}
		\label{fig:hyparm_3_16}
	\end{figure}
	\begin{figure}[tb]
		\centering
		\includegraphics[width=0.9\columnwidth]{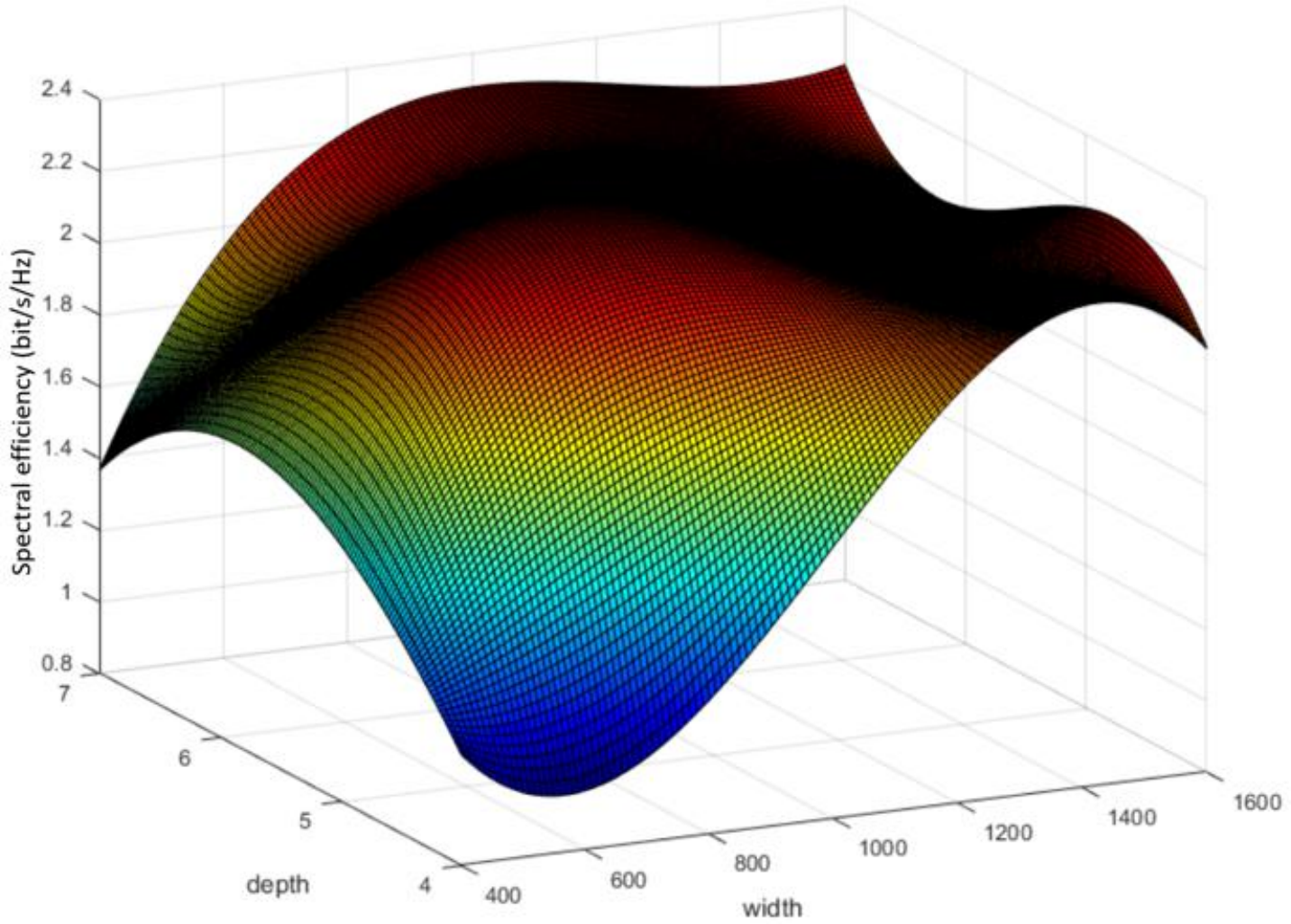}
		\caption{Spectral efficiency with various hyper parameters, where $B = 3$ and the number of devices per a cell = 24.}
		\label{fig:hyparm_3_24}
	\end{figure}

	
	Fig.~\ref{fig:110p} describes the effect of the transmit power constraint factor, $\lambda_p$, which is less sensitive than $\lambda_{if}$, because $P_{max}$ is 0.25 W. Similar to the case of $\lambda_{if}$, DNN may ignore the power constraint if $\lambda_{p}$ is not high enough. With a very small $\lambda_{p}$, $\eta$ can be increased but cannot maintain the constraint. DNN adopts the transmit power constraints appropriately where $\lambda_{p}$ is over 10. Unlike $\lambda_{if}$, a larger $\lambda_{p}$ does not has a problem. Even when $\lambda_{p}$ is 200, the performance of spectral efficiency does not change. It is because there is no D2D transmitter, which is over the $P_{max}$ after enough training.

	\Figure[!t]()[width=1.0\textwidth]{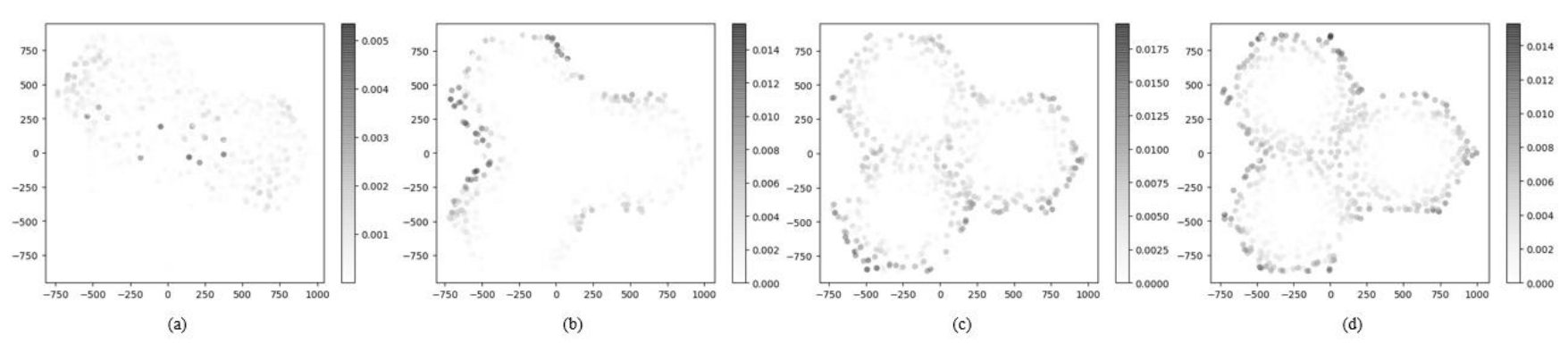}{Cell power distribution during the learning process, where $Q_{max}=-150$ dBW and $B = 3$ with various iterations; (a) 0, (b) 1K, (c) 10K, and (d) 100K. \label{fig:3process}}
	
	\Figure[!t]()[width=1.0\textwidth]{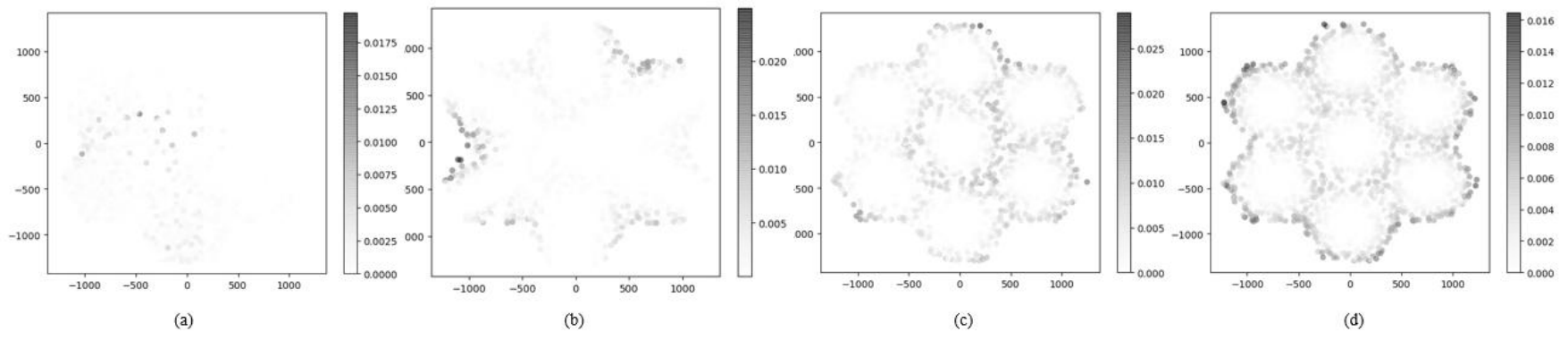}{Cell power distribution during the learning process, where $Q_{max}=-150$ dBW and $B = 7$ with various iterations; (a) 0, (b) 1K, (c) 10K, and (d) 100K. \label{fig:7process}}
	
	\Figure[!t]()[width=1.0\textwidth]{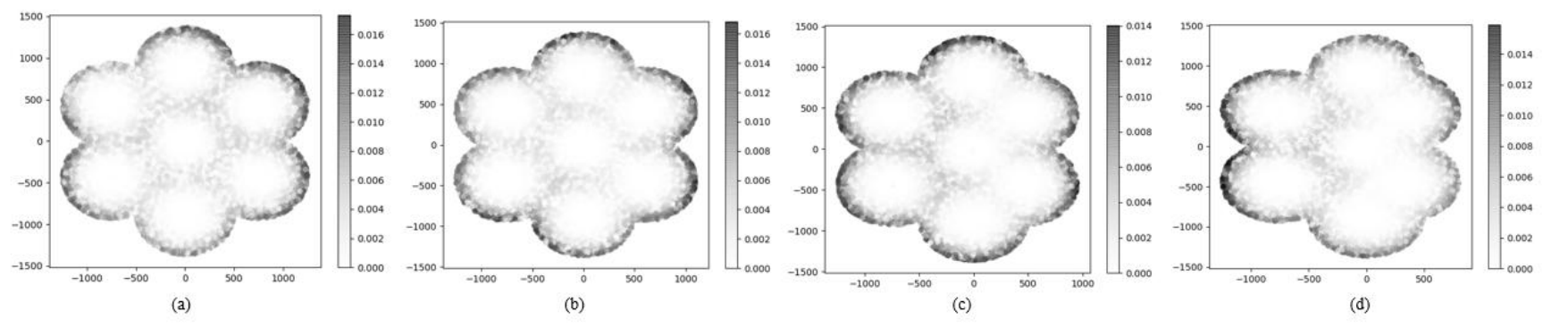}{Power distribution with distorted and non-hexagonal cell architecture by shifting the two right cell to left; (a) Non-shifted, (b) 20\%, (c) 40\%, and (d) 60\%. \label{fig:distorted}}

	Fig.~\ref{fig:pw_compare} compares the Iterative Approximated Distributed Rate Maximization 
	Problem with Interference Constraint (IADRMPIC) in~\cite{ref:d2d_conv} and the proposed scheme with various $P_{max}$ and $Q_{max}$. With the four cases of different $Q_{max}$, the proposed scheme has similar throughput to the IADRMPIC. Note that the purpose of the proposed scheme is to achieve similar throughput without any involvement of other nodes. It shows that DPADIC can achieve a meaningful throughput via a prediction method with deep learning.

	Fig.~\ref{fig:scala} shows the scalability of the proposed scheme. Scalability with various numbers of devices is important to a system because DPADIC uses pre-trained deep learning model. We compare two deep learning models which are trained with 8 pairs and 12 pairs, respectively. The learning model also considers the constraint of eNB interference factor, $\lambda_{if}$. The models have been tested for various numbers of devices: ranging from 2 to 24 pairs. Throughput decrease as the number of devices increases because of the effect of interference.
	
	Note that there is no meaningful difference between two pre-trained models. The deep learning model is trained to achieve that the eNB interference constraint in any distributions, so the policy from the deep learning is set conservatively. It means that there is a room for additional devices to meet the eNB interference constraint. According to this experiment, the pre-trained model can show valid performance for sufficiently diverse cases of the numbers of devices.

	Figs.~\ref{fig:hyparm_3_16} and~\ref{fig:hyparm_3_24} show that $T$ with various hyper parameter cases, where training with 16 devices and 24 devices respectively and $B = 3$. Depth means that the number of layers and width is the number of perceptrons in a layer. According to these experiments, both depth and width are important to achieve enough performance of deep learning. Note that the case of 24 devices requires more hyper parameters than those of the case of 16 devices. It means that the case of more devices is regarded as a more complex problem to solve. For scalability, it is advantageous to set higher hyper parameters. Optimizing hyper parameter is another challenging issue for most deep learning schemes~\cite{ref:hp1, ref:hp2}. However, the proposed scheme, it does not focus on optimizing hyper parameters. Also, the experimental results show that the range of valid hyper parameters is large enough. Therefore, an additional optimizing hyper parameter algorithm is not required. The reason for the low association between hyper parameters and spectral efficiency is that the large amount of data can prevent overfitting. Overfitting is a phenomenon where performance is rather poor when the size of the neural network is too large for the number of data. In this system, the data can be generated by simulation, so it is hard to have the overfitting problem.


	Fig.~\ref{fig:3process} and~\ref{fig:7process} show visualized training results for each cell environment respectively. Because of the interference constraints to eNB, the D2D power allocations are more distributed in a cell edge area. With 100k iterations, it can get almost converged results. These results indicate DNN divides the compartments for power allocation to maximize throughput. It allocates fractionally transmit power by very slight subdividing. In particular, it is remarkable that the transmit power of the cell in the edge area increase. This implies that D2D links with the proposed method can be helpful to improve throughput of cell edge users. The signals of cell edge users can be combined or multi-hopped by D2D communication. Furthermore, DPADIC can be derived in a distributed way, which means that the performance enhancement for the cell edge users can be conducted without eNB involvement.

	In Fig.~\ref{fig:distorted}, power distributions with distorted and non-hexagonal cell architecture where $Q_{max}=-150$ dBW and $B=7$ are depicted. To show that the proposed scheme can work independently from the architecture of cells, distorted cell architectures are simulated by shifting two right cells to left. Deep learning is a mapping function of the location and the transmit power to maximize cell throughput. Even if the distribution of the cell changes, the mapping ability of the deep learning does not decrease.

	\section{Conclusion}
	We propose a distributed power allocation scheme for D2D links underlaying a cellular system. We describe the models that the D2D devices work autonomously. Then, the sum of the results of decisions at each device can achieve near-optimal spectral efficiency of the related result.
	It can be expressed that the D2D devices memorize the appropriate transmit power with location information to meet the near-optimal result. The proposed method also has another technical point that can be generalized. There are two features that can be adopted for not only wireless communication but also other optimization problems. The first feature is that it supports to solve general maximizing problems while maintaining specific constraints using deep learning. We show that it can be operated to optimize a problem while maintaining several constraints. Another feature is the distributed deep learning architecture. We solve the distributed power allocation problem for D2D links using this architecture, which can be applied to develop a centralized system into a distributed system. In the future, we will improve the proposed scheme for more complex system, which is difficult to cope with conventional schemes. 
	
	\section*{Acknowledgment}
	This work was supported by the research fund of Signal Intelligence Research Center supervised by the Defense Acquisition Program Administration and Agency for Defense Development of Korea.

\begin{IEEEbiography}[{\includegraphics[width=1in,height=1.25in,clip,keepaspectratio]{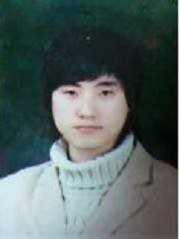}}]{Jeehyeong Kim}
		received the B.E. degree in Computer Science and Engineering from Hanyang University, South Korea, in 2015. He is currently pursuing the M.S.-leading-to-Ph.D. degree in Computer Science and Engineering at Hanyang University, South Korea. Since 2015, he has been with the Computer Science and Engineering, Hanyang University of Engineering, South Korea. His research interests include machine learning in cyber security and next wireless communications (email: manje111@gmail.com)
	\end{IEEEbiography}

\begin{IEEEbiography}[{\includegraphics[width=1in,height=1.25in,clip,keepaspectratio]{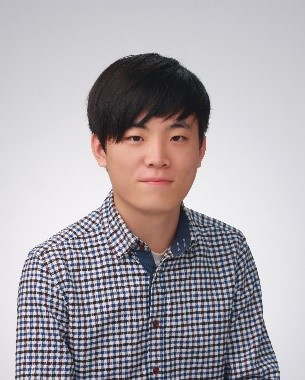}}]{Joohan Park}
		received the B.E. degree in Computer Science and Engineering from Hanyang University, South Korea, in 2017. He is currently pursuing the M.S.-leading-to-Ph.D. degree in Computer Science and Engineering at Hanyang University, South Korea. Since 2017, he has been with the Computer Science and Engineering, Hanyang University of Engineering, South Korea. His research interests include wireless power transfer, RF energy harvesting network, and simultaneous wireless information and power transfer (SWIPT).  (email: 1994pjh@hanyang.ac.kr)
	\end{IEEEbiography}

\begin{IEEEbiography}[{\includegraphics[width=1in,height=1.25in,clip,keepaspectratio]{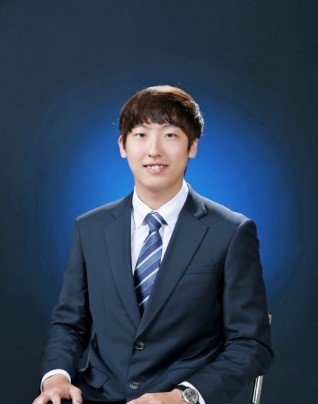}}]{Jaewon Noh}
		Noh received the B.E. degree in Computer Science and Engineering from Hanyang University, South Korea, in 2015. He is currently pursuing the M.S.-leading-to-Ph.D. degree in Computer Science and Engineering at Hanyang University, South Korea. Since 2015, he has been with the Computer Science and Engineering, Hanyang University of Engineering, South Korea. His research interests include security for wireless communication, wireless communication, and vehicular communication systems.  (email: wodnjs1451@hanyang.ac.kr)
	\end{IEEEbiography}
\begin{IEEEbiography}[{\includegraphics[width=1in,height=1.25in,clip,keepaspectratio]{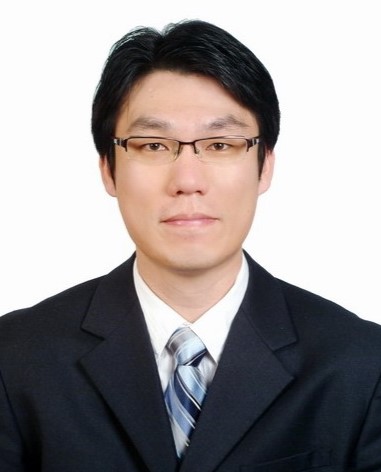}}]{Sunghyun Cho}
		received his B.S., M.S., and Ph.D. in Computer Science and Engineering from Hanyang University, Korea, in 1995, 1997, and 2001, respectively. From 2001 to 2006, he was with Samsung Advanced Institute of Technology, and with Telecommunication R\&D Center of Samsung Electronics, where he has been engaged in the design and standardization of MAC and network layers of WiBro/WiMAX and 4G-LTE systems. From 2006 to 2008, he was a Postdoctoral Visiting Scholar in the Department of Electrical Engineering, Stanford University. He is currently a Professor in the dept. of Computer Science and Engineering, Hanyang University. His primary research interests are 5th generation mobile communications, software defined networks, and vehicular communication systems. He is a member of the board of directors of the Institute of Electronics and Information Engineers (IEIE) and the Korean Institute of Communication Sciences (KICS).  (email: chopro@hanyang.ac.kr)
	\end{IEEEbiography}
\EOD


\begin{thebibliography}{1}
		
		\bibitem{ref:traffic1}
		D. Naboulsi, M. Fiore, S. Ribot, and R. Stanica, “Large-scale mobile traffic analysis: A survey,” \textit{IEEE Communication Surveys \& Tutorials}, vol. 18, no. 1, pp. 124–161, 1st Quart., 2016.
		
		\bibitem{ref:iot1}
		C. Suraci, S. Pizzi, A. Iera, and G. Araniti, “Enhance the protection of transmitted data in 5G D2D communications through the Social Internet of Things,” \textit{IEEE  International Symposium on Personal Indoor and Mobile Radio Communications}, vol. 2018-September, pp. 376–380, 2018.
		
		\bibitem{ref:iot2}
		J. Ni, X. Lin, and X. S. Shen, “Efficient and Secure Service-Oriented Authentication Supporting Network Slicing for 5G-Enabled IoT,” \textit{IEEE Journal on Selected Areas in Communications}, vol. 36, no. 3, pp. 644–657, 2018.
		
		\bibitem{ref:iot3}
		N. Jiang, Y. Deng, M. Condoluci, W. Guo, A. Nallanathan, and M. Dohler, “RACH Preamble Repetition in NB-IoT Network,” \textit{IEEE Communication Letters}, vol. 22, no. 6, pp. 1244–1247, 2018.
		
		\bibitem{ref:D2Dsurvey1}
		F. Jameel, Z. Hamid, F. Jabeen, S. Zeadally, and M. A. Javed, “A survey of device-to-device communications: Research issues and challenges,” \textit{IEEE Communication Surveys \& Tutorials}, vol. 20, no. 3, pp. 2133–2168, 2018.
	  
	    \bibitem{ref:D2Dsurvey2}
		M. K. Pedhadiya, R. K. Jha, and H. G. Bhatt, “Device to device communication: A survey,” \textit{Journal of Network and Computer Applications}, vol. 129, no. 1, pp. 71-89, Mar. 2019.
		
	    \bibitem{ref:D2Dsurvey3}
	    P. Gandotra, R. Kumar Jha, and S. Jain, “A survey on device-to-device (D2D) communication: Architecture and security issues,” \textit{Journal of Network and Computer Applications}, vol. 78, no. 1, pp. 9–29, Oct. 2017.
		
		\bibitem{ref:D2Dsurvey4}
		N. Cheng, et al, “Performance analysis of vehicular device-to-device underlay communication,” \textit{IEEE Transactions on Vehicular Technology}, Vol. 66, no. 6, pp. 5409-5421, 2017.

		\bibitem{ref:D2D_5G}
		S. A. A. Shah, E. Ahmed, M. Imran, and S. Zeadally, “5G for Vehicular Communications,” \textit{IEEE Communication Magazine}, vol. 56, no. 1, pp. 111–117, 2018.
		
        \bibitem{ref:drl_survey}
        N. C. Luong et al., “Applications of Deep Reinforcement Learning in Communications and Networking: A Survey,” \textit{IEEE Communication Surveys \& Tutorials}, vol. PP, no. c, pp. 1–1, 2019.
        
        \bibitem{ref:ex_dl_iot1}
		N. Jiang, Y. Deng, A. Nallanathan, and J. A. Chambers, “Reinforcement Learning for Real-Time Optimization in NB-IoT Networks,” \textit{IEEE Journal on Selected Areas in Communications}, vol. 37, no. 6, pp. 1424–1440, 2019.
		
		\bibitem{ref:ex_dl_iot2}
        N. Jiang, Y. Deng, O. Simeone, and A. Nallanathan, “Cooperative Deep Reinforcement Learning for Multiple-group NB-IoT Networks Optimization,” \textit{IEEE International Conference on Acoustics, Speech, and Signal Processing}y, Brighton, UK, pp. 8424–8428, May, 2019.


        \bibitem{ref:ex_dl_iot3}
        J. Tang, D. Sun, S. Liu, and J. L. Gaudiot, “Enabling Deep Learning on IoT Devices,” \textit{Computers}, vol. 50, no. 10, pp. 92–96, 2017.
    
        
        \bibitem{ref:ex_dl_iot4}
        H. Li, K. Ota, and M. Dong, “Learning IoT in Edge: Deep Learning for the Internet of Things with Edge Computing,” \textit{IEEE Networks}, vol. 32, no. 1, pp. 96–101, 2018.

		
		\bibitem{ref:chip}
		H. Li, M. Bhargava, P. N. Whatmough, and H. S. P. Wong, “On-Chip memory technology design space explorations for mobile deep neural network accelerators,” \textit{Design Automatic Conference}, Las Vegas, Nevada, pp. 1-6, Jun. 2019.
		
		\bibitem{ref:light}
		Xiangyu Zhang, Xinyu Zhou, Mengxiao Lin, and Jian Sun, "ShuffleNet: An Extremely Efficient Convolutional Neural Network for Mobile Devices," \textit{IEEE Conference on Computer Vision and Pattern Recognition}, Salt Lake City, Utah, pp. 6848-6856, Jun. 2018.

		\bibitem{ref:rel1}
		H. Dai, Y. Huang, R. Zhao, J. Wang, and L. Yang, “Resource Optimization for Device-to-Device and Small Cell Uplink Communications Underlaying Cellular Networks,” \textit{IEEE Transactions on Vehicular Technology}, vol. 67, no. 2, pp. 1187–1201, 2018.
		
\bibitem{ref:rel2}
Y. Han, X. Tao, and X. Zhang, “Power allocation for device-to-device underlay communication with femtocell using stackelberg game,” \textit{IEEE Wireless Communications and Networking Conference}, Barcelona, Spain, pp. 1–6, Apr. 2018.


\bibitem{ref:karim}
J. Kim, N. Karim, and S. Cho, "An Interference Mitigation Scheme of Device-to-Device communication for Sensor Networks Underlying LTE-A." \textit{Sensors}, vol. 17, no. 5, pp. 1-18, May. 2017.


\bibitem{ref:rel3}
J. Huang, C. C. Xing, Y. Qian, and Z. J. Haas, “Resource Allocation for Multicell Device-to-Device Communications Underlaying 5G Networks: A Game-Theoretic Mechanism with Incomplete Information,” \textit{IEEE Transactions on Vehicular Technology}, vol. 67, no. 3, pp. 2557–2570, Oct, 2018.

\bibitem{ref:rel4}
G. Giambene and T. A. Khoa, “Efficiency and fairness in the resource allocation to device-to-device communications in LTE-A,” \textit{IEEE International Conference on Communications}, Kansas City, MO, pp. 1–6, May, 2018.

\bibitem{ref:rel5}
A. Zappone, B. Matthiesen, and E. A. Jorswieck, “Energy Efficiency in MIMO Underlay and Overlay Device-to-Device Communications and Cognitive Radio Systems,” \textit{IEEE Transactions on Signal Processing}, vol. 65, no. 4, pp. 1026–1041, Nov. 2017.

\bibitem{ref:rel6}
S. Gupta, R. Zhang, and L. Hanzo, “Energy Harvesting Aided Device-to-Device Communication Underlaying the Cellular Downlink,” \textit{IEEE Access}, vol. 5, pp. 7405–7413, Aug. 2016.


\bibitem{ref:autol}
A. Asheralieva and Y. Miyanaga, “An autonomous learning-based algorithm for joint channel and power level selection by D2D pairs in heterogeneous cellular networks,” \textit{IEEE Transactions on Communications}, vol. 64, no. 9, pp. 3996–4012, Sep. 2016.






\bibitem{ref:d2d_conv}
A. Abrardo and M. Moretti, "Distributed Power Allocation for D2D communication Underlaying/Overlaying OFDMA Cellular Networks," \textit{IEEE Transactions on Wireless Communications}, vol. 16, no. 3, pp. 1466-1479, Dec. 2016.
		
		
\bibitem{ref:rel7} 3GPP TR 36.885 "Technical Specification Gropu Radio Access Network; Study on LTE-based V2X Services; Rel. 14," 3GPP Technical Report, Jun. 2016.

\bibitem{ref:rel8} R. Molina-Masegosa, and J. Gozalvez, "LTE-V for sidelink 5G V2X vehicular communications: A new 5G technology for short-range vehicle-to-everything communications," \textit{IEEE Vehicular Technology Magazine}, vol. 12, no. 4, pp. 30-39, Dec. 2017.

\bibitem{ref:rel9} J. Mei, K. Zheng, L. Zhao, Y. Teng, and X. Wang, “A Latency and Reliability Guaranteed Resource Allocation Scheme for LTE V2V Communication Systems,” 
\textit{IEEE Transactions on Wireless Communications}, vol. 17, no. 6, pp. 3850–3860, Mar. 2018.


\bibitem{ref:rel10} F. Abbas, P. Fan, and Z. Khan, "A Novel Low-Latency V2V Resource Allocation Scheme Based on Cellular V2X Communications," \textit{IEEE Transactions on Intelligent Transportation Systems}, vol. 20, no. 6, pp. 2185 - 2197, Jun, 2018.

\bibitem{ref:rel10-1}
Y. Liu, H. Yu, S. Xie, and Y. Zhang, “Deep Reinforcement Learning for Offloading and Resource Allocation in Vehicle Edge Computing and Networks,” \textit{IEEE Transactions on Vehicular Technology}, vol. PP, no. c, pp. 1–1, 2019.


\bibitem{ref:rel11} B. Soret, M.G. Sarret, I.Z. Kovacs, F. J. Martin-Vega, G. Berardinelli, and N. H. Mahmood, "Radio resource management for V2V discovery," \textit{IEEE Vehicular Technology Conference}, Sydney, Australia, pp. 1-6. Jun. 2017.

\bibitem{ref:rel12} H. Ye and G. Y. Li, “Deep Reinforcement Learning for Resource Allocation in V2V Communications,” \textit{IEEE International Conference on Communications}, Kansas City, MO, pp. 1–6, May, 2018.
	


\bibitem{ref:dl1}
Y. LeCun,  Y. Bengio, and G. Hinton, "Deep learning," \textit{Nature}, vol. 521, no. 7553, pp. 436-444, May. 2014.

	
	\bibitem{ref:wdl1}
	H. Sun, X. Chen, Q. Shi, M. Hong, X. Fu, and N. D. Sidiropoulos, “Learning to optimize: Training deep neural networks for wireless resource management,” \textit{IEEE International Workshop on Signal Processing Advances in Wireless Communications}, Sapporo, Japan, pp. 1–6, Jul. 2017.	
	
	\bibitem{ref:wmmse}
	Q. Shi, M. Razaviyayn, Z.-Q. Luo, and C. He, “An iteratively weighted MMSE approach to distributed sum-utility maximization for a MIMO interfering broadcast channel,” \textit{IEEE Transactions on Signal Processing}, vol. 59, no. 9, pp. 4331–4340, Apr. 2011.
	
	\bibitem{ref:wdl2}
	Y. He, C. Liang, F. Yu, N. Zhao, and H. Yin, “Optimization of cache-enabled opportunistic interference  alignment wireless networks: A big data deep reinforcement learning approach,” \textit{IEEE International Conference on Communications}, Paris, France, pp. 1-6, Jun. 2017.
	
	\bibitem{ref:wdl3}	
	G. Gui, H. Huang, Y. Song, and H. Sari, “Deep Learning for an Effective Nonorthogonal Multiple Access Scheme,” \textit{IEEE Transactions on Vehicular Technology}, vol. 67, no. 9, pp. 8440–8450, 2018.	
	
	\bibitem{ref:intlink}
	W. Xu, H. Zhou, H. Wu, F. Lyu, N. Cheng, and X. Shen, “Intelligent Link Adaptation in 802.11 Vehicular Networks: Challenges and Solutions,”  \textit{IEEE Communications Standards Magazine}, vol. 3, no. 1, pp. 12–18, Mar. 2019.
	
	\bibitem{ref:wdl4}	
	W. Lee, M. Kim, and D. Cho, “Transmit Power Control Using Deep Neural Network for Underlay Device-to-Device Communication,” \textit{IEEE Wireless Communication Letter}, vol. PP, no. c, p. 1, 2018.
	
	\bibitem{ref:T-DNN}	
	P. De Kerret, D. Gesbert, and M. Filippone, “Team deep neural networks for interference channels,” \textit{IEEE International Conference on communication Workshops}, Kansas City, MO, pp. 1–6, May. 2018.
	
	\bibitem{ref:resnet}
    Kaiming He, Xiangyu Zhang, Shaoqing Ren and Jian Sun, "Deep Residual Learning for Image Recognition," \textit{IEEE Conference on Computer Vision and Pattern Recognition}, Las Vevas, Nevada, pp. 770-778, Jun. 2016. 
	
	\bibitem{ref:xavier}
	X. Glorot and Y. Bengio, "Understanding the difficulty of training deep feedforward neural networks," \textit{Proceedings of the Thirteenth International Conference on Artificial Intelligence and Statistics}, Sardinia, Italy, pp. 249-256, May. 2010.
	
	\bibitem{ref:batch_norm}
	S. loffe and C. Szegedy, "Batch Normalization: Accelerating Deep Network Training by Reducing Internal Covariate Shift," \textit{Proceedings of the Thirteenth International Conference on Artificial Intelligence and Statistics}, San Deigo, CA, pp.448-456, May. 2015.
	
	\bibitem{ref:adam}
	D. Kingma and J. Ba, "Adam: A Method for Stochastic Optimization," \textit{International Conference for Learning Representations}, San Diego, CA, pp. 1-15, May. 2015.
	
	\bibitem{ref:v2v_modeling}
	C. Huang, A. F. Molisch, R. He, R. Wang, P. Tang and Z. Zhong, "Machine-Learning-Based Data Processing Techniques for Vehicle-to-Vehicle Channel Modeling," \textit{IEEE Communications Magazine}, vol. 57, no. 11, pp. 109-115, Nov. 2019.
	
	\bibitem{ref:winner}
	IST-WINNER D1.1.2 P. Kyösti, et al., "WINNER II Channel Models", ver 1.1, Sep. 2007. Available: https://www.istwinner.org/WINNER2
	
	\bibitem{ref:tensorflow}
	M. Abadi, et al. "Tensorflow: Large-scale machine learning on heterogeneous distributed systems," \textit{arXiv:1603.04467v2}, pp. 1-19, Mar. 2016; Available: https://arxiv.org/abs/1603.04467v2
	
	
	\bibitem{ref:hp1}
	M. U. Yaseen, A. Anjum, O. Rana, and N. Antonopoulos, “Deep learning hyper-parameter optimization for video analytics in clouds,” \textit{IEEE Transactions on Systems, Man, and Cybernetics: Systems}, vol. 49, no. 1, pp. 253–264, 2019.
	
	\bibitem{ref:hp2}
	S. R. Young, D. C. Rose, T. P. Karnowski, S. H. Lim, and R. M. Patton, “Optimizing deep learning hyper-parameters through an evolutionary algorithm,” \textit{Proceedings of the Workshop on Machine Learning in High-Performance Computing Environments}, Austin, TX, pp. 1-5, Nov. 2015	
	
		
		
		
		
		
		
		
		
		
		
		
	
		
		
		
		
		
		
		
		
		
		
		
		
		
		
		
		
		
		
		
		
		
		
		
		
		
		
		
	\end{thebibliography}
\end{document}